# Towards a possible discovery of dark photons with a multi-cathode counter.

© 2022 A. Kopylov[1]*,  I. Orekhov[1], V. Petukhov[1]

[1]Institute for Nuclear Research of Russian Academy of Sciences,

Moscow, Russia

*E-mail: kopylov@inr.ru

**Abstract** - The main points of experiment on the search of dark photons using multy-cathode counter are outlined. The evidence base to prove that the observed effect is really from dark photons is formulated. First data on diurnal variations are presented. The possible interpretation of the data is given for the case if they are expanded in future measurements.

1. INTRODUCTION.

The aim of our experiment PHELEX (PHoton-ELectron EXperiment) is to search for Dark Matter. The only empirical evidence for dark matter to-day comes from astrophysical observations. So we know that the dark matter accounts for about 84 % of the matter content of the Universe [1] (Planck 2018 results) but we don't know why it is so and what is a nature of dark matter? This is the first enigma. We know also that dark matter is distributed in galaxy by a spherical halo while baryonic matter – in the plane of galaxy. This constitutes the second enigma. Our future studies should give the answers to these questions. The most popular candidate for dark matter to-day is WIMPs – weakly interacting massive particles because some theoretical justification for this can be formulated. But however compelling are theoretical ideas, they are just like that: compelling but may be not true. So we should look for dark matter everywhere, where only the experiment is capable to shed a light on this question. We describe here our experiment aimed at the observation of dark photons – hypothetical Spin-1 Bosons with a rest mass as an alternative to standard electrodynamics. They were suggested in 80-ies [2-4]. The configuration of the fields of dark photons can be very complicated as it was shown recently in [5]. We use a specially developed for this purpose technique – multi-cathode counter [6] and measure the counting rate of single electrons emitted from a metallic cathode as a result of some conversion of dark photons at the surface of a cathode. The net result is found as a difference of count rates measured in two different configurations of electric field: in the first one we measure the signal plus background from the tracks of ionized particles at the edges of the counter while in second configuration only the background is measured. We make measurements run by run and get a final result by averaging the difference of count rates for many runs. The measurements are performed around the clock daily. We record not only the magnitude of the voltage on the output of charge-sensitive preamplifier but also the moments of time of each 100 ns interval. Each day we obtain about 1 TB of data. These data are analyzed off-line. We find the diurnal variations of the count rates of single electrons in solar frame and in stellar frame.

## 2. RESULTS AND THEIR POSSIBLE INTERPRETATION IF THE DATA ARE EXPANDED.

In the search for dark matter it is important not only to register the effect but also to prove that the measured effect is really from dark matter. Our method is unique in the way that it really can yield evidence for this. This is because the multi-cathode counter has directionality. Figure 1 shows it.

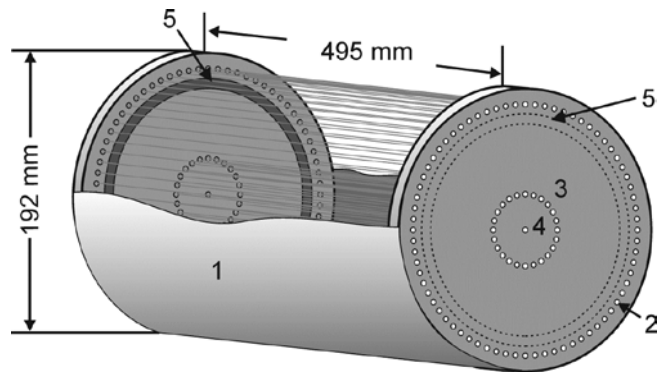

Figure 1. The simplified schematic of multi-cathode counter. 1 – metallic cathode, 2 – second cathode, 3 – third cathode, 4 – anode, 5 – focusing rings.

If the vector of $E$-field of dark photons is along the axis of the counter the effect should be zero. If this vector is perpendicular to the axes of the counter, the effect should be maximal. This is explained by the expression

$$P = 2\alpha^2\chi^2\rho_{CDM}A_{dish} \qquad (1)$$

Here: *P* – is a power absorbed by a metallic cathode during conversion of dark photon to usual photon, $\alpha^2 = cos\theta^2$ where $\theta$ is the angle between vector of *E*-field and a one normal to the surface of the element of the cathode, $\chi$ is kinetic mixing parameter quantifying the process of mixing dark photons with the real ones, $\rho_{CDM}$ is the energy density of dark matter and $A_{dish}$ is the surface of the cathode. Suppose the detector is placed at a site at certain geographical latitude. When the Earth rotates one should observe diurnal variations. The curve of the variation depends upon geographical latitude of the site, the orientation of the counter (vertical, horizontal West-East, horizontal North-South) and the angle $\varkappa$ between vector of *E*-field and the axes of the Earth due to varying factor $\alpha^2$. The calculation shows [7] that one obtains very different curves with the exclusion that they are similar for different geographical latitudes if the detector is oriented horizontal West-East what actually has a trivial geometrical explanation. But the experiment should confirm this fact if the signal is really from dark photons with a certain polarization in solar or stellar frame. Another thing is that the curves should be symmetrical relative a moment of time 12-00 if for the moment 00-00 one takes the one when the vector of E-field is in the plane of a meridian where the detector is placed[1] [8]. The substantial moment here is also that internal surface of the cathode should be polished to provide specular reflection. For the rough surface one should not observe diurnal variations due to this effect. One should also note that one solar day is 24 hours while one stellar day is 23 h 56 min 4 s. The frame where one observes curves of diurnal variations with 12-hours symmetry is a true frame where dark photons have a certain polarization. The real curves observed in experiment can be different from the ones calculated using $\alpha^2 = cos^2\theta$ in expression (1) [9] because we do not know in details the physics of the conversion of dark photons at the surface of the metal. One should note also that the target in this case is free electrons of a degenerate electron gas in a metal with certain polycrystalline structure. To summarise, we can list the validity conditions that the observed effect is really from dark photons:

1. The curve of diurnal variation must be symmetrical relative some moment in solar or stellar frame
2. The curves must be different for counters placed in sites with different geographical latitudes if only the counter is not oriented horizontally from West to East.
3. The curves must be different for counters with different orientations (vertical, horizontal W-E, N-S)
4. The effect should not be observed with counters with a rough internal surface of a cathode.

---

[1] More correct, this is correct only for solar frame where two moments of symmetry exist: 00-00 and 12-00 where vector of E field is in the plane of meridian. For stellar frame the points are 00-00 and 11-58 because stellar day is shorter: 23 h 56m.

If these conditions are fulfilled one can find the moment of time when the vector of E-field was in the plane of a meridian of the site where detector was placed. Comparing the curves of diurnal variations with the calculated one the angle ϰ between vector of E-field and the axes of the Earth can be found. In this way the vector of polarization of dark photons can be determined in solar or stellar frame. If any contradiction is observed between the calculated curve and the one obtained in measurements this will be a hint that something has been omitted in our considerations. The explanation of this can be found comparing diurnal curves for counters with different orientation and/or placed at sites with different geographical latitudes. All of this forms a basis for our future study.

Figure 2 shows the calculated curves of diurnal variations calculated for the counter placed in a steel shielded cabinet at the basement of a building in Troitsk, Moscow.

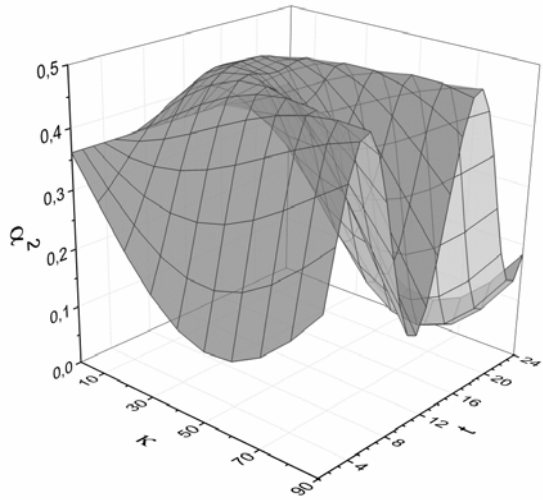

Figure 2. The curves of diurnal variations for the counter placed at the ground floor of a building at Troitsk, Moscow. Time $t$ is in hours starting from the moment when vector of $E$-field is in the plane of a meridian of the site where detector is placed.

The calculations have been done for real position of the counter in this cabinet. The counter was horizontally oriented with its axes at 23 degrees relative meridian. Because it was not oriented right along meridian one observes some asymmetry in the curves of diurnal variations relative the moment 12-00.

Figure 3 shows the diurnal variations obtained with a very limited statistics of data for this counter. One can see that the error bars here are large.

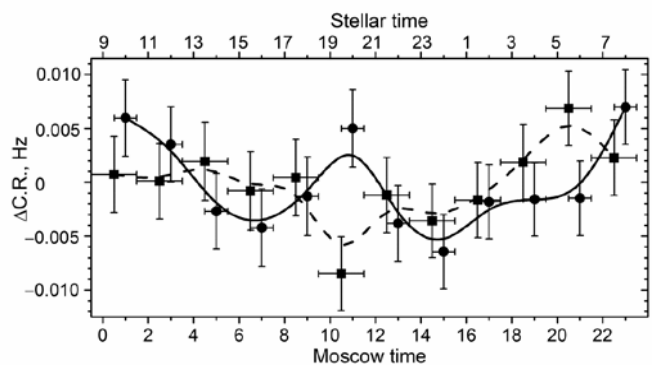

Figure 3. The results of measurements of the diurnal variations. Solid line and lower scale – solar frame, dashed line and upper scale – stellar frame. $\Delta$C.R. – the difference of the count rates measured in $1^{st}$ and $2^{nd}$ configurations in counts per second.

This figure is presented just to show what can be interpretation if the data are expanded and error bars are small. First, we should look for some curve symmetrical relative some moment of time in solar or stellar frame. The curves which do not exhibit this feature can be discarded as the false ones. Second, this moment of time should be at 12-00 in solar frame (11-58 in stellar frame) if for the moment 00-00 the one when the vector of *E*-field is in the plane of meridian of the site where detector is placed. This should be observed for the counter oriented vertically or horizontally strictly along meridian or along parallel of the Earth. So from the comparison what should be and what are really observed one can find at what time in solar or stellar frame this happens to be. And third, we should look whether the measured curves of the diurnal variations agree with the calculated ones. If yes, the expression (1) is valid. If not, maybe we have omitted something in our considerations so some corrections are needed to this expression.

### 3. CONCLUSIONS.

The method developed by us for the search of dark photons has shown its efficiency. The result obtained [10] has been included recently in compilation of data by Particle Data Group [11]. The essential feature of this technique is that it can present evidence that the observed effect is really from dark photons with a certain polarization in solar or stellar frame. The focus of our present study is to expand data collected to search for diurnal variations of the signal in our detector.

We appreciate very much the substantial support from the Ministry of Science and Higher Education of Russian Federation within the "Instrument Base Renewal Program" in the framework of the State project "Science".